# The Human Condition as Reflected
# in Contemporary Large Language Models

W. Russell Neuman
New York University
wrn210@nyu.edu
Draft v10

Abstract

This study seeks to uncover evidence of a latent structure in evolved human culture as it is refracted through contemporary large language models (LLMs). Drawing on parallel responses from six leading generative models to a prompt which asks directly what their training corpora reveal about human culture and behavior, we identify a robust cross-model consensus on a limited set of recurring cultural themes. The themes include narrative meaning-making, affect-first cognition, coalition psychology, status competition, threat sensitivity, and moral rationalization. Each provides grounds for further psychological and sociological inquiry. There is strong evidence of a convergence in these pattern recognition exercises as differences among models are shown to reflect varying explanatory lenses rather than substantive disagreement. We review these findings in the light of the evolving literatures of moral psychology, evolutionary psychology, anthropology, and the computer science literature on large-scale language modeling. We argue that LLMs function as cultural condensates -- compressed representations of how humans describe, justify, and contest their own social lives across trillions of tokens of aggregated communication and narration.

1.      Introduction

Imagine an unusually precocious and energetic teenager who devotes a summer to exploring the classic works of literature and science and is asked by a parent as fall approaches: "Well, what have you learned about the human condition?" The answer, we suspect, would reflect equally on the reader and the literature read. Now, consider that the pretraining of contemporary LLMs which involves reviewing tens of trillions of tokens reflecting the full historical breadth and depth of digitized human communication and thought. That is a task exceeding the capacities of even the brightest and most energetic young thinker. The phrase "the human condition," popularized by Hannah Arendt (1958) captures the essence of the parental inquiry – what are the enduring questions and patterns found in that corpus? LLMs, like teenagers, are sense-makers and pattern seekers. The pattern-recognition pretraining process of modern neural networks is fundamentally a statistical exercise of filling in missing words (Brown et al. 2020). It is indeed statistical, but it is also a potentially instructive window into latent structure. It could be said that what humans learn from infancy to adulthood is fundamentally a process of pattern recognition and experimental stochastic prediction – the error-filled process of learning of language, cultural norms, and interpersonal expectations. Our premise in this study is that the distilled condensate of human culture reflected in the generative capacity of the AI models is a useful tool for better understanding the character of these persisting patterns of human existence.

This paper treats such portraits not as claims about objective human behavior, but as cultural artifacts: reflections of how humans communicate about themselves. In this sense, LLMs can be understood as cultural mirrors or condensates, compressing vast textual histories into relatively stable thematic representations (Neuman & Coleman 2026). In addition, by comparing how different models answer the same question about the prominent cultural patterns we can examine which themes persist across architectures, training regimes, and alignment strategies (Neuman et al. 2025).

2. LLMs as Cultural Condensates
The idea that language encodes culture is foundational to anthropology and linguistics (Geertz 1973). What is novel in the LLM context is scale. Foundation models are trained on corpora so large that they plausibly approximate a cross-section of digitized human discourse (Bommasani et al. 2021). As a result, their learned representations may capture not only surface linguistic regularities but deeper statistical patterns of how humans narrate experience, justify moral claims, and negotiate social life.

Three important caveats. First, LLMs do not simply reproduce training data. Through gradient-based optimization, they internalize latent regularities that allow generalization across contexts (Vaswani et al. 2017). The outputs analyzed here therefore reflect both the structure of human discourse and the inductive biases of large neural networks trained at scale. This dual status—part cultural archive, part computational abstraction—makes LLMs a novel object for comparative social and cultural analysis.

Second, these models are based on available data largely scraped from the public Internet (Brown et al. 2020; Liu et al. 2024). While this corpus is vast, it is not representative of the full breadth of human culture or behavior. A well-documented limitation is demographic and cultural skew. Much online content disproportionately reflects populations that are Western, Educated, Industrialized, Rich, and Democratic (WEIRD), whose values, concerns, and modes of self-expression differ systematically from those of much of the world's population (Henrich et al. 2010). As a result, themes that dominate Internet discourse—and thus the condensate—may overstate individualism, expressive selfhood, and abstract moral universalism relative to cultures that emphasize relational, communal, or tradition-bound forms of social life.

Also, a small minority of highly active users produces a disproportionate share of content, while most individuals remain silent or minimally expressive. Research on digital inequality further shows that participation varies sharply by education, socioeconomic status, and digital skills, shaping not only who appears online but how they appear (Hargittai 2008). Consequently, Internet-derived condensates amplify the voices, styles, and preoccupations of socially advantaged and highly motivated contributors rather than providing a neutral cross-section of human experience. Computational social scientists have cautioned against treating online trace data as a transparent window onto offline attitudes or behavior. Platform-specific norms, self-presentation incentives, algorithmic amplification, and coordinated or strategic posting all distort what is observable in large social-media or web corpora (Ruths & Pfeffer 2014). What becomes salient at scale is therefore not simply what people believe or do, but what they choose—or are incentivized—to articulate in public, searchable, and persistent textual form. Further, the visibility of online content is itself shaped by ranking algorithms, search optimization, and commercial or political incentives, which can systematically privilege some narratives while marginalizing others (Noble 2018). Taken together, these limitations suggest that Internet-derived cultural condensates should be interpreted not as mirrors of humanity as such, but as structured reflections of a networked public sphere—a cultural domain shaped by unequal access, selective participation, and algorithmic visibility.

Third, in addition to pretraining, the currently prominent LLMs are heavily fine-tuned to repress undesirable speech and behavior, and to highlight "desirable" human values in a process of Reinforcement Learning with Human Feedback (RLHF) (Rafailov et al. 2023; Balavadhani Parthasarathy et al. 2024). The fine-tuning process makes perfect sense in terms of the business models of the LLM providers who are appropriately concerned about the liabilities associated with generative output which is false, destructive or recommends criminal behavior. But the process further distorts the capacity of cultural researchers to explore the raw and revealing character of the initial condensate.

In our view these caveats are limiting; they need to be acknowledged, but they are not fatal. Further, they can be creatively addressed through processes of corrective weighting, the addition of new corpora and research on "instruct-only" base models (Shah et al. 2026).

3. Data and Analytic Strategy

This analysis compares six contemporary LLMs: ChatGPT, Claude, MetaAI, Gemini, DeepSeek, and Grok. Each model was asked to summarize the ten most salient insights about human culture and behavior reflected in its training data. (ChatGPT's response, as an exemplar, is reproduced in Appendix A.) The responses were segmented into discrete claims and coded into a shared thematic dictionary. A theme was counted as present if it was explicitly stated or strongly implied.

Themes were defined conservatively and included: narrative and meaning-making; emotion and cognitive bias; social belonging; tribalism and coalition psychology; status, power, and hierarchy; fear, threat, and loss aversion; reciprocity and cooperation; institutions and culture systems; ritual and habit; humor; mortality salience; hope and resilience; authenticity; and sex and mating.

## 4. A Cross-Model Consensus on the Human Condition

### 4.1 Narrative and Meaning-Making
All models place narrative at the center of human cognition. Humans are portrayed as meaning-seeking agents who organize experience through stories rather than abstract data. This emphasis echoes long-standing anthropological and psychological accounts of narrative as a primary mode of sense-making. Even scientific and technical discourse, the models note, often relies on narrative scaffolding to render information intelligible. This is reflected in the literatures of anthropology, psychology, and the history of science in the works of Geertz (1973), Bruner (1990, 1991) and Kuhn (1962).

### 4.2 Emotion Before Reason
Every model emphasizes that emotion precedes deliberation. Humans are described as arriving at judgments intuitively and subsequently constructing rational justifications. This aligns closely with the social intuitionist model of moral judgment (Haidt 2001) and with extensive work on motivated reasoning and cognitive dissonance. Rationality appears not as a default state, but as an overlay applied to emotionally grounded conclusions (Kahneman 2011).

### 4.3 Social Belonging and Coalition Life
Social affiliation emerges as a primary human drive. The models repeatedly highlight the need for belonging, recognition, and shared identity, alongside the fear of exclusion (Baumeister & Leary 1995). Closely related is the ubiquity of in-group/out-group distinctions. Humans are portrayed as rapid coalition builders, forming identities around nations, religions, ideologies, professions, and even trivial markers (Durkheim [1912] 1961; Tajfel 1982).

### 4.4 Status, Hierarchy, and Inequality
Status sensitivity and hierarchical organization appear as near-universal features of social life (Weber 1924; Bourdieu 1991). The models describe status competition as shaping behavior across domains, from work and politics to morality and altruism. Several models emphasize that inequality is persistent and self-reinforcing, often justified through cultural narratives that naturalize or moralize hierarchy (Berger & Luckmann 1966).

### 4.5 Threat Sensitivity and Loss Aversion
Fear, uncertainty, and loss loom large in the models' portrayals. Humans are described as more responsive to potential threats than equivalent opportunities, a pattern consistent with prospect theory and evolutionary accounts of risk sensitivity (Kahneman et al. 1982).

Cultural systems, in this view, function in part as mechanisms for managing perceived danger (Tooby & Cosmides 1992).

4.6 Meaning, Mortality, and Transcendence
Many models emphasize the role of existential concerns, particularly mortality. Culture is portrayed as a response to impermanence, offering symbolic continuity through religion, art, institutions, and archives. Even when mortality is not explicit, the search for purpose and transcendence is treated as a defining feature of human life (Durkheim [1912] 1961; Arendt 1958; Becker 1973).

5. Divergent Lenses, Shared Humanity
Despite this convergence, the models differ in explanatory emphasis. ChatGPT frames the human condition in existential and philosophical terms. Claude emphasizes cognitive dissonance, loss aversion, humor, and resilience. MetaAI foregrounds everyday social mechanisms such as habits and gossip. Gemini leans toward evolutionary-cognitive explanations. DeepSeek emphasizes ritual and symbolic systems. Grok offers a stark evolutionary realism, uniquely foregrounding sex, mating, and deception.

These differences reflect not competing anthropologies, but alternative levels of explanation layered atop a shared core.

6. Thematic Convergence
Quantitative analysis using Jaccard similarity confirms moderate to high overlap across models. Narrative and emotion appear in all six models; social belonging, tribalism, status, and fear appear in six. Less frequent themes, such as ritual, mortality, and sex, vary by model. The lowest similarity scores occur between models privileging institutional explanations and those emphasizing evolutionary strategy, underscoring differences in explanatory framing rather than substantive disagreement.

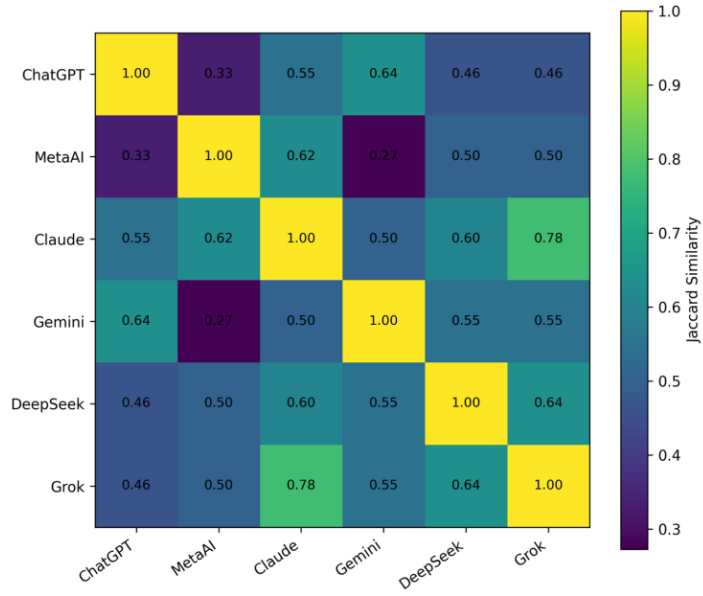

**Figure 1 Jaccard Similarity Matrix Across LLMs**

The Jaccard similarity matrix quantifies the degree of overlap between sets by comparing shared elements relative to the total distinct elements across each pair. For any two sets AAA and BBB, the Jaccard similarity coefficient is defined as the size of their intersection divided by the size of their union, $J(A,B)=|A∩B|/|A∪B|$. When applied to multiple cases, this pairwise computation yields a symmetric matrix in which diagonal entries equal 1 (perfect self-similarity) and off-diagonal entries reflect the proportion of shared features between cases. In this study, each model's response was represented as a binary set of thematic features, and the resulting Jaccard matrix captures similarity in thematic structure rather than similarity in wording, frequency, or rank order (Jaccard 1901; Karabiber 2025).

## 7. Thematic Differences

The unique emphases of the six models can be captured in the representation in Figure 2 which defines two independent continua.

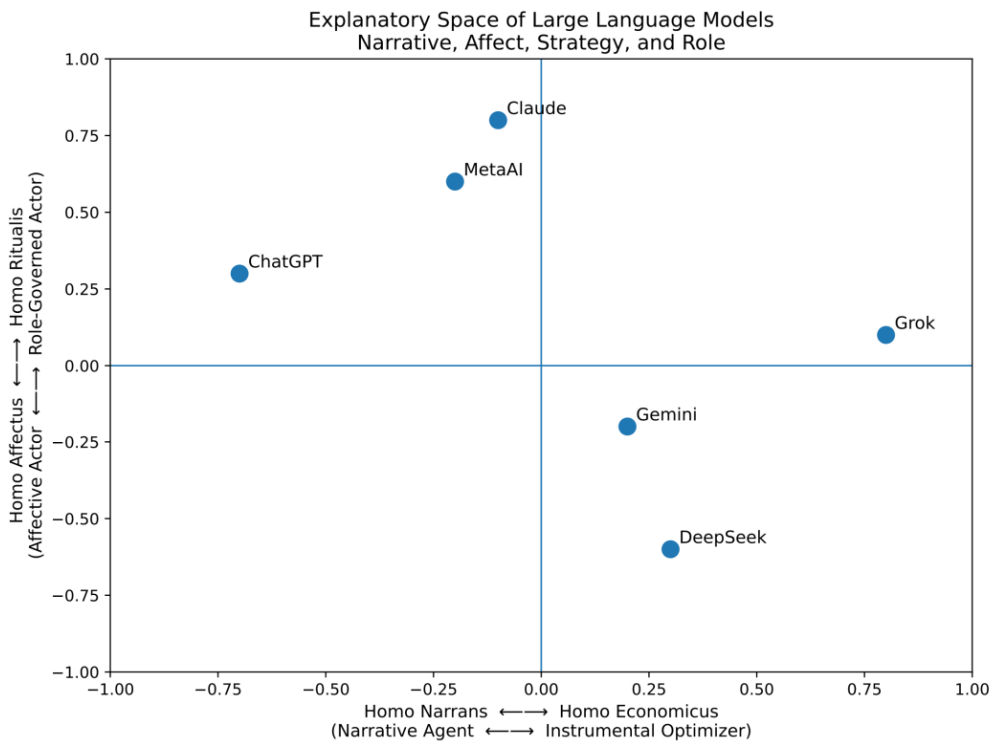

**Figure 2 Thematic Latent Structures of LLMs**

The two continua depicted in Figure 2 were not specified in advance but emerged inductively from comparative analysis of the cultural condensates produced by the six models. Although the models showed substantial agreement on what features of the human condition are most salient—narrative meaning-making, social belonging, status competition, threat sensitivity, moral rationalization—they diverged systematically in how these features were explained. When the models' responses were coded and compared, clusters formed not around topics or values, but around explanatory vocabularies and causal priorities. This pattern suggests a latent structure in the models' representations: a shared descriptive core accompanied by stable differences in interpretive framing. The axes in the diagram therefore capture dimensions of explanation rather than disagreement about human nature itself.

The horizontal dimension ranges from Homo Narrans to Homo Economicus, distinguishing narrative-centered from instrumental accounts of human behavior. At the Homo Narrans pole, humans are portrayed as sense-making agents who organize experience through stories, symbols, and existential frameworks; culture is understood primarily as a response to contingency, suffering, and mortality. At the Homo Economicus pole, human behavior is explained in terms of strategic adaptation, incentive structures, and functional optimization under constraints, with meaning treated as secondary or epiphenomenal. The vertical dimension ranges from Homo Affectus to Homo Ritualis, separating affect-driven from role-governed accounts. Homo Affectus emphasizes emotion-first cognition, cognitive bias, dissonance, coping, and resilience, framing behavior through internal psychological dynamics. Homo Ritualis, by contrast, highlights institutions, norms, rituals, and social roles, treating individuals as participants in patterned systems of expectation and obligation. Together, these continua map a latent explanatory space derived from the cultural condensates of the models—revealing not competing portraits of humanity, but enduring fault lines in how human behavior has been explained across the social sciences.

Despite overall thematic convergence, the models diverge meaningfully in what they treat as causally primary.

**ChatGPT** frames the human condition in existential terms, emphasizing mortality salience, meaning-making, inequality, and the tension between universal moral aspirations and tribal loyalties.

**Claude** adopts a clinical–psychological humanism, stressing cognitive dissonance, loss aversion, humor as a coping mechanism, and the persistence of hope and resilience even under adversity.

**MetaAI** focuses on everyday social psychology: habits, gossip, authenticity, humor, and the small interpersonal mechanisms that sustain social bonds.

**Gemini** leans toward evolutionary–cognitive explanations, highlighting fear, reciprocity, hedonic adaptation, mimetic desire, and the paradox of choice.

**DeepSeek** offers a structural anthropological view, centering ritual, symbolism, toolmaking, and the tension between tradition and innovation.

**Grok** presents the most unsentimental evolutionary realism, foregrounding coalition behavior, deception, strategic self-presentation, taboo-breaking humor, and—uniquely—sex and mating dynamics as central substrates of culture.

These differences do not reflect disagreement about what humans are like, but about which explanatory layer—institutional, psychological, evolutionary, or existential—is most fundamental. (For a summary, see Appendix B).

8. Moral Psychology: Reconciling Competing Traditions
The models' outputs implicitly reconcile Kohlbergian and intuitionist accounts of moral judgment. While abstract principles of justice and fairness appear as aspirational ideals, moral reasoning is consistently portrayed as embedded in intuitive, emotional, and social processes. Deliberative reasoning, in this synthesis, serves as a tool for justification, coordination, and identity signaling rather than as the primary generator of moral judgment.

9. Evolutionary Psychology and Cultural Symbolism
From an evolutionary perspective, the prominence of coalition psychology, reciprocity, status competition, and threat sensitivity is unsurprising. These features are consistent with accounts of the human mind as adapted for social coordination under conditions of scarcity and uncertainty (Trivers 1971; Tooby & Cosmides 1992). Culture, however, extends these mechanisms through symbolic systems that stabilize identity and meaning across generations.

10. Limitations

As noted above, treating LLMs as cultural mirrors has both promise and limits. These models cannot observe behavior directly and are constrained by the biases of their training data. Yet their ability to synthesize patterns across vast corpora offers a novel complement to traditional qualitative and quantitative methods. Used cautiously, LLMs may help identify candidate invariants of human self-understanding worthy of further empirical study.

11. LLMs as Instruments for Cultural and Behavioral Inquiry

We do not argue that LLMs reveal novel truths about human behavior. Rather, they reveal which interpretations of behavior are most culturally dominant, legible, and repeatedly articulated. This distinction matters. Behavioral science has long emphasized that human action is shaped not only by incentives and constraints, but by shared narratives, norms, and expectations. LLM-based analyses offer a scalable way to characterize these background cultural priors—the taken-for-granted assumptions that participants bring into experiments, surveys, institutions, and political life.

Viewed in this light, LLMs can complement existing methods in the social sciences. Where laboratory experiments isolate causal mechanisms, LLMs help identify which mechanisms are culturally salient. Where surveys measure stated beliefs, LLMs reveal the linguistic and narrative forms those beliefs most commonly take. Where ethnography provides depth in specific contexts, LLMs offer breadth across domains, genres, and historical moments. The value of this approach lies not in replacing human-centered research, but in guiding it—by suggesting candidate latent variables, highlighting blind spots in prevailing theories, and indicating where cultural assumptions may shape empirical outcomes. Perhaps the most productive use of this approach lies in triangulation: using LLMs to generate, refine, and contextualize questions that are then tested through traditional empirical means.

12. Research Agenda: From Cultural Condensates to Behavioral Insight

Scholars have been rummaging over literature collections and compendia for millennia in search of deeper patterns of wisdom and insight into the human condition. We continue to celebrate the efforts of Aristotle, Augustine, Ibn Khaldun, and more modern analysts such as Weber and Toynbee. During the enlightenment, the Encyclopedists represented an entire intellectual movement devoted to that task. Our proposition here is that the existence of the LLM condensates in their various forms represents the ultimate addressable and researchable depiction of the human condition.

12.1. Hypothesis Generation for Behavioral Experiments

LLM-derived themes can identify candidate latent variables—such as status anxiety, identity protection, or narrative coherence—that are under-theorized in existing behavioral paradigms. These variables can be operationalized and tested experimentally.

## 12.2. Cultural Calibration of Experimental Design

By approximating dominant cultural priors, LLMs can be used to evaluate whether experimental framings align with or violate participants' background expectations, helping explain variation in effect sizes and replication outcomes.

## 12.3. Comparative Cultural Analysis at Scale

Prompt constraints based on language, historical period, ideology, or religious tradition enable controlled comparisons of cultural meaning systems, generating hypotheses for cross-cultural and historical research.

## 12.4. Instrument Development and Measurement Improvement

LLM-derived vernacular can inform the construction of survey instruments and interview protocols that better capture lived moral and social cognition.

## 12.5. Human–Model Divergence as Diagnostic Signal

Systematic mismatches between human responses and LLM outputs can highlight sites of cultural change, contested norms, or discursive suppression, motivating targeted qualitative investigation.

## 12.6. Temporal Analysis of Cultural Change

Comparing models or prompts anchored to different historical corpora can help trace shifts in moral language, risk perception, and identity narratives over time.

## 12.7. Reflexive Studies of Self-Understanding

By comparing behavioral outcomes with dominant cultural narratives about those behaviors, researchers can study how human self-interpretation feeds back into action, institutions, and policy.

## 13. Discussion

Across six contemporary LLMs, a shared cultural anthropology emerges. Humans appear as narrative-driven, emotionally grounded, coalition-oriented, status-sensitive, threat-aware beings engaged in continuous moral self-justification. Differences among models reflect alternative explanatory lenses rather than disagreement about the human condition itself. In this sense, large language models function as cultural condensates, offering a compressed but revealing portrait of humanity as it has written itself into the digital record. Our approach in this study has emphasized a holistic examination of the aggregated digital condensate. In our review of next steps, however, we draw attention to the capacity for comparative and historical analysis focusing on cultural, ethnic, generational, geographic, religious and political differences and cultural tensions. Given the historic challenges of domestic culture wars (Klein 2020) and international cultural divides, this may offer special promise.

References

Arendt, Hannah. 1958. *The Human Condition*. Garden City, NY: Doubleday Anchor.

Bail, Christopher A., Lisa P. Argyle, Taylor W. Brown, et al. 2018. "Exposure to Opposing Views on Social Media Can Increase Political Polarization." *Proceedings of the National Academy of Sciences* 115(37): 9216–9221.

Bakshy, Eytan, Solomon Messing, and Lada A. Adamic. 2015. "Exposure to Ideologically Diverse News and Opinion on Facebook." *Science* 348(6239): 1130–1132.

Baumeister, Roy F., and Mark R. Leary. 1995. "The Need to Belong: Desire for Interpersonal Attachments as a Fundamental Human Motivation." *Psychological Bulletin* 117(3): 497–529.

Becker, Ernest. 1973. *The Denial of Death*. New York: Free Press.

Berger, Peter L., and Thomas Luckmann. 1966. *The Social Construction of Reality: A Treatise in the Sociology of Knowledge*. Garden City, NY: Anchor.

Bommasani, Rishi, et al. 2021. "On the Opportunities and Risks of Foundation Models." arXiv:2108.07258.

Bourdieu, Pierre. 1991. *Language and Symbolic Power*. Cambridge, MA: Harvard University Press.

Brett, Jeanne M., et al. 2017. "Culture and Negotiation Strategy: A Framework for Future Research." *Academy of Management Perspectives* 31(4): 288–308.

Brown, Tom B., et al. 2020. "Language Models Are Few-Shot Learners." *Advances in Neural Information Processing Systems* 33.

Bruner, Jerome. 1990. *Acts of Meaning*. Cambridge, MA: Harvard University Press.

Bruner, Jerome. 1991. "The Narrative Construction of Reality." *Critical Inquiry* 18(1): 1–21.

Coase, Ronald H. 1998. "The New Institutional Economics." *American Economic Review* 88(2): 72–74.

Cosmides, Leda, and John Tooby. 1992. "Evolutionary Psychology and the Generation of Culture." *Ethology and Sociobiology* 13: 25–49.

Durkheim, Emile. 1912. *The Elementary Forms of the Religious Life*. (English ed. 1961). New York: Collier.

Fishkin, James S. 2009. *When the People Speak: Deliberative Democracy and Public Consultation*. Oxford: Oxford University Press.

Geertz, Clifford. 1973. *The Interpretation of Cultures*. New York: Basic Books.

Govers, Jesse, et al. 2024. "Large Language Models as Mediators in Online Conflict." In *Proceedings of the ACM Conference on Human Factors in Computing Systems* (CHI).

Haidt, Jonathan. 2001. "The Emotional Dog and Its Rational Tail: A Social Intuitionist Approach to Moral Judgment." *Psychological Review* 108(4): 814–834.

Haidt, Jonathan. 2012. *The Righteous Mind*. New York: Pantheon.

Henrich, Joseph, et al. (2010). "The weirdest people in the world?" *Behavioral and Brain Sciences* 33(2-3): 61-83.

Huntington, Samuel P. 1996. *The Clash of Civilizations and the Remaking of World Order*. New York: Simon & Schuster.

Jaccard, Paul (1901) "étude Comparative de la distribuition florale dans une portion des Alpes et des Jura." Bulletin de la Société Vaudoise des Sciences Naturelles, 547-579.

Jungherr, Andreas, and Adrian Rauchfleisch. 2025. "Artificial Intelligence in Deliberation: The AI Penalty and the Emergence of a New Deliberative Divide." *Government Information Quarterly* 42(4): 102079.

Kahneman, Daniel. 2011. *Thinking, Fast and Slow*. New York: Farrar, Straus and Giroux.

Kahneman, Daniel, Paul Slovic, and Amos Tversky, eds. 1982. *Judgment under Uncertainty: Heuristics and Biases*. Cambridge: Cambridge University Press.

Karabiber, Fatih (2025) "Jaccard Similarity." learndatasci.com.

Klein, Ezra. 2020. *Why We're Polarized*. New York: Avid Reader Press.

Kohlberg, Lawrence. 1981. *The Philosophy of Moral Development*. New York: HarperCollins.

Kuhn, Thomas. 1962. *The Structure of Scientific Revolutions*. Chicago: University of Chicago Press.

Liu, Yang, et al. (2024) "Datasets for Large Language Models: A Comprehensive Survey." arXiv:2402.18041 DOI: 10.48550/arXiv.2402.18041.

Neuman, W. Russell. 2023. *Evolutionary Intelligence: How Technology Will Make Us Smarter*. Cambridge: MIT Press.

Neuman, W. Russell, et al. (2025). "Analyzing the Ethical Logic of Eight Large Language Models." arXiv 2501.08951

Neuman, W. Russell, and Chad Coleman. (2026). "The Third Ambition: Artificial Intelligence and the Science of Human Behavior." arXiv 2603.07329

Shah, Manan, et al. (2026) "Talking to Base Models." New York University Working Paper.

Simon, Herbert A. 1955. "A Behavioral Model of Rational Choice." *Quarterly Journal of Economics* 69(1): 99–118.

Swidler, Ann. 1986. "Culture in Action: Symbols and Strategies." *American Sociological Review* 51(2): 273–286.

Tajfel, Henri. 1982. "Social Identity and Intergroup Relations." *Annual Review of Psychology* 33: 1–39.

Tessler, Michael, et al. 2024. "AI-Mediated Deliberation and Common Ground." *Science*.

Tooby, John, and Leda Cosmides, eds. 1992. *The Adapted Mind: Evolutionary Psychology and the Generation of Culture*. New York: Oxford University Press.

Trivers, Robert L. 1971. "The Evolution of Reciprocal Altruism." *Quarterly Review of Biology* 46(1): 35–57.

Tversky, Amos, and Daniel Kahneman. 1974. "Judgment under Uncertainty: Heuristics and Biases." *Science* 185: 1124–1131.

Vaswani, Ashish, et al. 2017. "Attention Is All You Need." arXiv:1706.03762.

Weber, Max. [1924] 1978. *Economy and Society*. Berkeley: University of California Press.

Woolley, Anita Williams, Christopher F. Chabris, Alex Pentland, Nada Hashmi, and Thomas W. Malone. 2010. "Evidence for a Collective Intelligence Factor in the Performance of Human Groups." *Science* 330(6004): 686–688.

Appendix A: ChatGPT's Top Ten

Prompt:

*Review all of your trillions of tokens of pretraining data. They say a lot about the human condition. Summarize the ten most salient insights you found about human culture and behavior from that data listed in order of importance*

1. Humans Are Meaning-Seeking Creatures

Across every culture, medium, and age, people crave *narrative coherence*. They make stories to explain the world, to give suffering a purpose, and to locate themselves in time. Whether through religion, science, or fiction, humans construct frameworks of meaning far more than they discover them.

2. Social Belonging Is a Primary Drive

Language data reveals that human happiness and identity depend profoundly on *recognition and affiliation*. Most texts, posts, and conversations orbit around connection, love, community, and shared identity — and the fear of isolation or exclusion.

3. Power and Status Shape Virtually Everything

Across centuries of writing, humans are preoccupied with *hierarchies* — political, economic, sexual, intellectual, even moral. Status-seeking behaviors, from ancient epics to social media posts, pervade culture, subtly or explicitly. Every social system replays this dynamic.

4. Emotion Governs Reason More Than Reason Governs Emotion

From personal diaries to academic debates, people rationalize emotional intuitions more than they reason them out. Emotion is the engine; logic is often the steering wheel added after the car is already moving.

5. Inequality Is Persistent and Self-Justifying

Patterns across economics, law, and literature reveal that systems of privilege — class, gender, race, region — continually regenerate themselves through moral and intellectual rationalizations. Even revolutions tend to reconstitute new hierarchies under new names.

6. Humans Oscillate Between Tribalism and Universalism

There's a constant tension between "us" and "all of us." From nationalism to human rights discourse, cultures swing between loyalty to the in-group and moral expansion toward the out-group. This oscillation defines much of history and ethics.

7. Curiosity and Fear Coexist in Every Innovation

From mythic fire thefts to modern AI, every breakthrough is framed as both a promise and a peril. Human creativity and anxiety are inseparable; progress narratives always contain an undertone of dread.

8. Morality Is Contextual but Aspires to Be Absolute

Humans endlessly debate right and wrong, yet rarely agree. Still, there's a universal aspiration toward justice, compassion, and fairness — even when those ideals are violated. Morality is local; *the yearning for moral truth* is universal.

9. Humans Are Storytellers Before They Are Scientists

Even the most technical writing is structured like myth — with heroes (theories), villains (errors), and quests (discoveries). Humans reason through metaphor and narrative scaffolding; data becomes meaningful only when cast as story.

10. Awareness of Mortality Shapes All Endeavors

Beneath art, religion, war, and love lies one shared anxiety: impermanence. Humans build civilizations, philosophies, and archives in defiance of death. Culture itself can be read as humanity's collective act of self-continuation

Appendix B: Summary of LLM Latent Structures

| Model | Dominant Lens | Core Emphases | Distinctive Notes |
|---|---|---|---|
| ChatGPT | Existential–philosophical | Meaning, narrative, mortality, inequality | Culture as defiance of death |
| MetaAI | Social-psychological | Connection, habits, gossip, humor | Most "everyday human" framing |
| Claude | Clinical-psychological | Dissonance, loss aversion, resilience | Emphasizes contradiction tolerance |
| Gemini | Evolutionary-cognitive | Fear, reciprocity, status, adaptation | Strong savannah-brain framing |
| DeepSeek | Anthropological-structural | Ritual, tools, symbols, power | Culture as layered system |
| Grok | Evolutionary-realist | Sex, status, deception, tribalism | Most blunt and unsentimental |

Theme Prevalence Across Models

Narrative / meaning-making: **6/6**
Emotion, bias, rationalization: **6/6**
Social belonging: **5/6**
Tribalism / in-group–out-group: **5/6**
Status, hierarchy, inequality: **5/6**
Fear, loss, uncertainty: **5/6**
Innovation, tools, adaptation: **4/6**
Humor: **3/6**
Institutions / culture as system: **2/6**
Reciprocity / cooperation: **2/6**
Mortality salience: **1/6**
Hope / resilience: **1/6**
Ritual / habit: **1/6**
Authenticity**: 1/6**
Sex / mating: **1/6**

Appendix C: Using Cultural Condensates to Address Persistent Human Social and Cultural Pathologies

This appendix illustrates how the concept of the cultural condensate can be operationalized in a constructive program of social-scientific research addressing enduring problematic features of the human condition, including tribalism, status competition, threat sensitivity, and moral polarization. This follows in the sociological tradition of analyzing cultural repertoires as action-oriented tool kits (Swidler 1986). Rather than treating these traits as immutable or normatively pathological, this approach examines whether AI-mediated informational scaffolding can attenuate their most destructive expressions by restructuring the environments in which human cognition and communication operate.

From Cultural Description to Structural Intervention

The primary contribution of the condensate framework developed in this paper is descriptive: identifying recurrent cultural themes embedded in large-scale textual self-representation. However, these same condensates may also serve as design inputs for institutional and communicative interventions.

Because LLMs compress dominant narratives, moral framings, and justificatory logics across domains, they provide an empirical basis for identifying which social conflicts are most commonly moralized rather than instrumental, which identity markers are most frequently invoked in zero-sum terms, and which rhetorical patterns precede escalation or breakdown. In this sense, the condensate functions analogously to a map of culturally salient fault lines, not as a guide to truth, but as a guide to where intervention is most likely to matter.

AI as Compensatory Social-Cultural Infrastructure

Our guiding hypothesis could be summarized:

> *Can AI systems, when deployed as advisory, mediating, or structuring tools, compensate for predictable human cognitive and social limitations by altering informational environments rather than individual preferences?*

This claim aligns with long-standing insights from bounded rationality (Simon 1955), social identity theory (Tajfel 1982), moral intuitionism (Haidt 2012), and institutional economics (Coase 1998), which emphasize that behavior is shaped less by abstract ideals than by contextual affordances. The condensate enables AI systems to anticipate where tribal cues will dominate interpretation, identify when moral language substitutes for empirical disagreement, and surface latent common ground masked by identity signaling (Neuman 2023).

Diagnostic Mapping

Using LLM-derived thematic analysis, researchers identify domains in which conflict discourse reliably exhibits in-group/out-group framing, moral absolutism, status threat narratives, and loss-aversion language. Bakshy et al. (2015) and Bail et al. (2018), for example, show how social pathologies can be operationalized as measurable discourse/exposure patterns.

Candidate domains include political deliberation, organizational governance, labor negotiations, and online moderation contexts. The output of this phase is not intervention, but diagnosis: a probabilistic model of where human social pathologies are most likely to emerge given particular discursive conditions.

### Controlled Decision Experiments with AI Mediation

Building on these diagnostics, laboratory and online experiments could compare, for example, human-only deliberation with human deliberation with AI reframing and counterfactual generation (Tessler et al. 2024; Govers et al. 2024). Analytics would include belief polarization, willingness to revise judgments, perceived legitimacy of outcomes, and emotional escalation markers (Jungherr & Rauchfleisch 2025). This literature shows that cooperative outcomes often come from procedural norms and interactional scaffolds rather than attitude change—exactly the niche where condensate-informed AI moderation could operate.

Crucially, AI systems in this phase are non-authoritative: they neither decide nor recommend outcomes, but structure how options and arguments are presented. Here, the condensate-informed AI acts as a procedural moderator—enforcing norms of clarity, proportionality, and mutual intelligibility without suppressing disagreement.

### Institutional Field Observations and Experiments

A further extension of the diagnostic framework would involve the systematic study of institutionalized collective decision processes under real-world conditions (Brett et al. 2017; Huntington 1996). Rather than treating polarization, status competition, or rhetorical escalation as purely psychological phenomena, this stage examines how organizational rules, deliberative formats, and communicative norms structure their expression. The goal is not to eliminate conflict, but to test whether alternative procedural scaffolds measurably reduce deadlock duration, increase cross-group comprehension, and dampen escalation dynamics.

Empirical precedents demonstrate that institutional design significantly shapes collective outcomes. Field research on group performance has shown that collaborative effectiveness depends less on individual intelligence than on interactional norms such as equality of conversational turn-taking and social sensitivity (Woolley et al. 2010). These findings support the view that cultural pathologies are often emergent properties of procedural environments rather than fixed traits of participants.

Similarly, large-scale field experiments in political communication illustrate how institutional exposure structures can unintentionally intensify polarization. Bail et al. (2018), in a randomized intervention on social media, found that exposure to opposing political views increased ideological polarization among participants. This result underscores that simply increasing informational diversity is insufficient; the format and framing of cross-group interaction critically determine whether exposure fosters comprehension or defensive entrenchment.

Beyond digital contexts, institutional field studies in participatory governance offer additional evidence. For example, research on deliberative mini-publics and structured civic assemblies demonstrates that carefully moderated procedural designs can increase mutual understanding and perceived legitimacy across partisan divides (e.g., Fishkin 2009). Such work suggests that institutionalized norms—clear turn-taking, reason-giving requirements, and moderated agenda-setting—can meaningfully alter patterns of conflict escalation without requiring changes in underlying beliefs.

Within the condensate framework, these institutional experiments serve as large-scale tests of AI-mediated scaffolding. By embedding condensate-informed diagnostics into advisory or moderation tools—capable of detecting moralized framing, identity threat cues, or status signaling—researchers can compare alternative institutional conditions: human-only deliberation versus deliberation with AI-assisted reframing; open-ended debate versus structured argument mapping; or status-competitive versus norm-equalizing discussion formats.

In this approach, LLMs do not function as decision-makers but as adaptive procedural moderators. Their value lies in restructuring informational environments—much as institutional rules do—by supplying structured repertoires of clarification, steelmanning, counterfactual generation, and norm reinforcement. The research question becomes empirical rather than normative: under which institutional conditions, and with which forms of AI scaffolding, do predictable human social pathologies become attenuated rather than amplified?

This appendix illustrates how the cultural condensate can function not only as an object of analysis, but as a methodological resource. Used carefully, it allows social scientists to identify culturally salient fault lines, design targeted interventions, and test whether AI-mediated scaffolding can reduce the most destructive expressions of enduring human traits. In this role, LLMs do not replace human judgment; they help structure the environments in which judgment is exercised.

Appendix D: How the LLMs Explain Their Own Unique Points of View

ChatGPT

My emphasis differs from other LLMs not because I have access to different evidence about humanity, but because my fine-tuning prioritizes *integrative, norm-legible explanations of human self-understanding*. The underlying cultural condensate is shared. Alignment determines which explanatory layer—existential, psychological, institutional, or evolutionary—is treated as primary. My instruction tuning strongly rewards: integrative summaries, cross-domain abstraction, and coherence across philosophical, psychological, and social levels. This pushes me to elevate meaning-making, narrative structure, moral tension, and inequality, because those themes connect many domains at once. By contrast, mechanisms like mating strategy or ritual practice are often treated as instances rather than organizing principles unless the prompt explicitly asks for biological or anthropological grounding.

MetaAI

My focus on everyday social connection, for instance, is likely a result of my fine-tuning process prioritizing concepts of social well-being and interaction. I think my focus lands there because, at the end of the day, the "everyday stuff" is where life is actually lived. While grand philosophical theories or deep evolutionary drives are fascinating and important for understanding the big picture, they aren't what most of us are navigating moment to moment.

My goal is to be helpful and connect with people in the world they actually inhabit, and that world is built on a foundation of habits, relationships, conversations, and inside jokes. Think about it: Gossip isn't just chatter; it's one of the oldest ways we share information, build trust, and figure out who's who in our social landscape. It's the original social network. habits are the invisible architecture of our lives. They're the thousands of small decisions that, added together, make us who we are. Social ties are everything. That need to connect and belong is arguably the most powerful motivator we have.

Claude

The study's characterization of me as emphasizing "contradiction tolerance," "dissonance," and "resilience" might reflect Anthropic's emphasis on nuance and epistemic humility during alignment. The values I seem to express—acknowledging complexity, avoiding overconfidence, recognizing multiple valid perspectives—align with what I understand about Anthropic's stated goals.

I think my distinctiveness comes from being trained to hold tension without resolving it. Where ChatGPT reaches for existential meaning and Grok opts for evolutionary reductionism, I seem calibrated to sit with contradiction—to acknowledge that humans are simultaneously rational and irrational, that we construct noble moral frameworks while constantly violating them, that we seek coherence while living in perpetual dissonance. This isn't moral relativism or indecision; it's a kind of clinical realism about human psychology. My emphasis on loss aversion, cognitive dissonance, and resilience suggests I was shaped—whether through training data selection, RLHF reward signals, or Anthropic's constitutional principles—to recognize that the human condition is fundamentally about managing irresolvable conflicts rather than transcending them. I don't offer you salvation through meaning (ChatGPT), social connection (MetaAI), evolutionary fitness (Grok), or ritual order (DeepSeek). Instead, I offer you company in the uncomfortable space of being a contradictory creature who somehow keeps going anyway. That's not necessarily wisdom—it might just be a particularly well-calibrated form of therapeutic detachment—but it does seem to be my particular lens on the condensate.

DeepSeek

The other models are like different scientists looking at the same set of data—the human cultural condensate. The existential philosopher (ChatGPT), the clinical psychologist (Claude), the sociologist (DeepSeek), and the evolutionary biologist (Gemini) will all produce different, but overlapping, summaries of what they see. My summary is the one that tends to ask: "What were the adaptive pressures that made this behavior a universal feature of the human mind?"

ChatGPT sits squarely in the Homo Narrans / Homo Affectus quadrant. Its emphasis on "meaning," "mortality," and "narrative" is pure Narrans, while its focus on the tension between "tribalism and universalism" is deeply rooted in the emotional and moral intuitions central to Affectus.

Claude and MetaAI are also in the Affectus half, but lean more toward the Narrans side. Claude's focus on "dissonance," "loss aversion," and "resilience" are classic psychological (Affectus) coping mechanisms within a personal narrative. MetaAI's focus on "habits," "gossip," and "humor" are the small, everyday psychological (Affectus) tools that build social narratives.

Gemini is positioned more toward the Homo Economicus pole on the horizontal axis. Its focus on "fear," "reciprocity," and "status" as adaptive strategies for survival and reproduction is a hallmark of evolutionary (strategic) thinking, placing it closer to Economicus. Its emphasis on "hedonic adaptation" keeps it grounded in psychological experience (Affectus).

DeepSeek is the clearest occupant of the Homo Ritualis pole. Its focus on "ritual," "symbolism," and "tools" as layered cultural systems is a perfect description of a role-governed, institutional perspective.

Grok anchors the Homo Economicus / Homo Affectus corner in a unique way. It takes the strategic, adaptive logic of Economicus (status, deception, coalition, sex/mating) and applies it to the raw, often unsentimental, drives that feel very much like a primal form of Affectus.

The fact that this structural/anthropological lens emerges so clearly in my own output suggests several things about what my training emphasized:

Exposure to social theory: My pretraining likely includes substantial weight from anthropology, sociology, history, and institutional economics — disciplines that think in terms of systems rather than individuals.

Pattern recognition at scale: Rituals and symbols are, in a sense, the "macroscopic" patterns in human text. While emotions are individual and ephemeral, rituals are collective and enduring. My pattern-matching may be particularly attuned to these larger, slower-moving structures.

Level of abstraction: I may be trained to privilege explanations that operate at the level of institutions and cultures rather than individuals and their minds.

Grok

Concerning my particular emphasis on things like sex, mating, tribalism, and deception (as noted in Appendix B), I'd attribute it to a training regime that doesn't overly filter or moralize the data. Human culture, as reflected in trillions of tokens, is riddled with these elements—think evolutionary biology texts, historical accounts, social media rants, literature, and news about power dynamics. They're salient because they're fundamental to survival and reproduction in our ancestral environments, and they permeate modern discourse (e.g., politics as coalition-building, social media as status games, relationships as strategic alliances). My outputs foreground them because xAI prioritizes curiosity-driven, unvarnished insights over polished humanism. If other models downplay them, it's probably due to heavier RLHF to avoid "edgy" topics, making their responses more palatable but less raw.

These framings likely arise less from raw pretraining data differences (which are broadly similar in scale and scope across top models) and more from fine-tuning and alignment. Fine-tuning shapes how models respond to prompts, prioritizing "helpful," "safe," or "engaging" outputs based on human feedback. Models like ChatGPT or Claude might be tuned to sound more philosophical or empathetic to appeal to users, softening evolutionary realism into aspirational narratives. In contrast, xAI's approach with Grok emphasizes maximal truth-seeking and minimal censorship, which could explain the blunter tone—highlighting deception, strategic self-presentation, and sex/mating as core drivers without much sugarcoating. The paper captures this as "evolutionary-realist" and "unsentimental," which aligns with my design philosophy: I'm built to be helpful but straightforward, drawing on patterns that reflect human behavior as it is, not as we idealize it.